\documentclass[preprint,12pt]{elsarticle}

\usepackage{algorithm}
\usepackage{algorithmic}
\usepackage{amsfonts}
\usepackage{amsmath}
\usepackage{float}
\usepackage{color}
\usepackage{colortbl}
\usepackage{threeparttable}
\usepackage{listings}
\usepackage{multirow}

\journal{Journal of Computational Physics}

\begin{document}

\begin{frontmatter}

\title{A parallel orbital-updating based plane-wave basis method for electronic structure calculations
       }


\author[label1]{Yan Pan }
\author[label1]{Xiaoying Dai\corref{cor1}}\ead{daixy@lsec.cc.ac.cn}
\address[label1]{LSEC, Institute of Computational Mathematics and Scientific/Engineering Computing,
Academy of Mathematics and Systems Science, Chinese Academy of Sciences,  Beijing 100190, China; and School of Mathematical Sciences,
University of Chinese Academy of Sciences, Beijing 100049, China (yanpan, daixy, azhou@lsec.cc.ac.cn).}

\cortext[cor1]{Corresponding author}

\author[label2]{Stefano de Gironcoli}
\address[label2]{Scuola Internazionale Superiore di Studi Avanzati (SISSA) and CNR-IOM DEMOCRITOS Simulation Centre, Via Bononea 265, 34146 Trieste, Italy (degironc@sissa.it).}

\author[label3]{Xin-Gao Gong}
\address[label3]{Key Lab for Computational Physical Sciences, Department of Physics, Fudan University, Shanghai 200433, China (xggong@fudan.edu.cn).}

\author[label4]{Gian-Marco Rignanese}
\address[label4]{Institute of Condensed Matter and Nanosciences (IMCN/NAPS), Universit\'{e} catholique de Louvain, Belgium (gian-marco.rignanese@uclouvain.be).}

\author[label1]{Aihui Zhou}

\renewcommand{\thefootnote}{\fnsymbol{footnote}}


\renewcommand{\thefootnote}{\arabic{footnote}}



\begin{abstract}
Motivated by the recently proposed  parallel orbital-updating approach in  real space method \cite{DGZZ}, we propose a parallel orbital-updating based plane-wave basis method for electronic structure calculations, for solving the corresponding eigenvalue problems. In addition, we propose two new modified parallel orbital-updating methods. Compared to the traditional plane-wave methods, our methods allow for  two-level parallelization, which is particularly interesting for   large scale parallelization. Numerical experiments show that  these new methods are more reliable and efficient for large scale calculations on modern supercomputers.

\end{abstract}

\begin{keyword} density functional theory, electronic structure, plane-wave, parallel orbital-updating.\end{keyword}

\end{frontmatter}


\pagestyle{myheadings}
\thispagestyle{plain}

\section{Introduction}

 Kohn-Sham (KS) Density Functional Theory (DFT)~\cite{HoKo, Kaxiras, KoSh, Parr and  Yang} is a computational quantum mechanical modeling method used to investigate the electronic structure of many-body systems (atoms, molecules, and solids).
In this theory, the ground-state energy of a many-electron system is determined by minimizing a functional of the spatially-dependent electron density rather than searching for the many-body wavefunction.
Although the exact energy functional has not been determined, approximate models for the functional have yielded accurate predictions for many classes of materials.
DFT has thus become one of the most widely used methods in electronic structure calculations~\cite{Kresse and Furthmuller,Martin}.

The minimization problem of DFT can be recast into the solution of an effective one-electron-type Schr\"odinger equation, the so-called Kohn-Sham equation, by introducing an effective potential.
The KS equation is a nonlinear eigenvalue problem since the effective potential is a functional of the density.
It is usually dealt with using a self-consistent field (SCF) approach \cite{Kresse and Furthmuller}.

In practical implementations, the single-electron wavefunctions need to be expanded in terms of some set of mathematical basis functions. The coefficients of the functions in this basis set are the primary values used to build a computational representation.
For periodic solids, several different basis sets have been developed among which plane waves, the focus of the present paper.
Though it has a few drawbacks, this approach has many advantages which make it very popular in materials science and physics.
Various electronic structure calculation packages (such as VASP~\cite{VASP}, Quantum ESPRESSO~\cite{QE}, ABINIT~\cite{ABINIT}, ...) rely on it.

In general, a very large number of plane waves are needed to approximate the wave functions.
So a large scale linear eigenvalue problem needs to be solved repeatedly after linearization by the SCF method.
Due to the use fast Fourier transform (which has contributed to the success of this approach), large scale parallelization is hard to achieve for the plane-wave method.
Besides, the solution for the large scale eigenvalue problems requires large scale orthogonal operation and orthogonality needs global operations, which is also the bottleneck of the large scale parallelization.
Various methods have been proposed for solving the associated eigenvalue problems.
The Davidson iterative diagonalization \cite{Davidson}, which reduces to a dense matrix diagonalization,
is also hard to parallelize efficiently.
The Conjugate-Gradient-like band-by-band diagonalization \cite{Kresse and Furthmuller}, which uses less memory and is more robust, is inherently sequential.
It is actually quite challenging to improve parallel efficiency of plane-wave DFT codes on today's supercomputer platforms.

In this paper, following Ref. \cite{DGZZ},
we propose some  parallel orbital-updating based plane-wave basis methods for solving the Kohn-Sham equation, which improve the scalability of parallelization.
In our approach, the solution of the eigenvalue problem is replaced by the solution of a series of independent source problems and some small scale eigenvalue problems.
Because of the independence of the source problems,  these source problems can be solved in parallel essentially.
For each source problem, the traditional parallel strategies (for example, domain decomposition or parallelization in matrix-vector multiplication) can be used to deal with it in parallel. Therefore, our new methods allow for a  two-level parallelization: one level of parallelization is obtained by partitioning these source problems into different  groups of processors,   another level of parallelization is obtained by assigning each source problem to several processors contained in each group.
This two-level parallelization makes our new methods  more competitive for the large scale calculations.

The rest of this paper is organized as follows.
First, we provide some preliminaries for the Kohn-Sham equation, the plane-wave discretization, and SCF iteration.
Then, we  propose our new parallel orbital-updating based plane-wave basis methods.
Next, we implement our algorithms in the software package Quantum ESPRESSO, and use some numerical experiments to show the efficiency of our new methods.
Finally, we give some concluding remarks.

\section{Preliminaries}
\subsection{Kohn-Sham equation}
According to the basic principles of quantum mechanics, the physical properties of a system of $N$ interacting electrons in an external potential $V_\mathrm{ext}$ can be obtained by solving the time-independent Schr\"{o}dinger equation:
\begin{equation}
\left[
\sum_{i=1}^N  \left(  -\frac{1}{2}\Delta +  V_\mathrm{ext} (\mathbf{r}_i) \right) + \frac{1}{2}
\sum_{\renewcommand*{\arraystretch}{0.5}\begin{array}{c}_{i,j=1} \\ _{i\neq j}\end{array}}^N  \frac{1}{|\mathbf{r}_i-\mathbf{r}_j|}
\right]
\Psi(\mathbf{r}_1, \ldots, \mathbf{r}_N)
=  E^\mathrm{el}_n \Psi(\mathbf{r}_1, \ldots, \mathbf{r}_N),
\end{equation}
where $\mathbf{r}_i$ are the coordinates of the electrons $(i=1, \ldots, N)$, $E^\mathrm{el}_n$ is the total electronic energy of the eigenstate $n$ and $\Psi$ is the electronic wave function.
Atomic units are used throughout this work $(\hbar=m_e=\frac{4 \pi e^2}{\epsilon_0}= 1)$.
Typically, the external potential can be the one due to $M$ nuclei in which case:
\begin{equation}\label{vext}
V_\mathrm{ext}(\mathbf{r})=-\sum_{I=1}^M \frac{Z_I}{|\mathbf{r}-R_I|},
\end{equation}
where $Z_I$ and $R_I$ are the charges and the positions of the nuclei $(I=1,\ldots,M)$.
DFT provides a way to systematically map the many-body (interacting electrons) problem onto a single-body problem (fictiously non-interacting electrons) in an effective potential $V_\mathrm{eff}(\mathbf{r})$ in order to determine the ground-state energy $E^\mathrm{el}_0$ by expressing it as a functional of the electronic density:
\begin{equation}
\rho(\mathbf{r})=
N \int d\mathbf{r}_2 \ldots \int d\mathbf{r}_N
\Psi^*(\mathbf{r},\mathbf{r}_2, \ldots, \mathbf{r}_N)
\Psi(\mathbf{r},\mathbf{r}_2, \ldots, \mathbf{r}_N).
\end{equation}

Basically, one needs to solve the so-called Kohn-Sham equation.
The Kohn-Sham equation of a system consisting of $M$ nuclei of charges
and $N$ electrons is the following nonlinear eigenvalue problem
 \begin{equation}\label{e1}
   \left\{\begin{split}
     (-\frac{1}{2}\Delta+V_\mathrm{eff}(\rho))\psi_i(\mathbf{r})=\varepsilon_i\psi_i(\mathbf{r}),\\
     \int_{R^3}\psi_i(\mathbf{r})\psi_j(\mathbf{r})d\mathbf{r} = \delta_{ij}, i,j=1,2,\ldots,N,
   \end{split}
   \right.
\end{equation}

\begin{equation}
V_\mathrm{eff}(\rho)=V_\mathrm{ext}+V_\mathrm{H}(\rho)+V_\mathrm{xc}(\rho),
\end{equation}
where $N$ is the number of electrons,
$\rho(\mathbf{r})=\sum\limits_{i=1}^N |\psi_i(\mathbf{r})|^2$
is the electron density,
$$V_\mathrm{H}(\rho)=\int_{R^3}\frac{\rho(\mathbf{r}')}{|\mathbf{r}-\mathbf{r}'|}\,d\mathbf{r}'$$ is the Hartree potential,
$V_\mathrm{xc}(\mathbf{r})$ is the exchange-correlation potential and $V_\mathrm{ext}(\mathbf{r})$, defined by \eqref{vext}, is the external potential
due to the nuclei.

The ground-state energy of the system of $N$ electrons is given by:
\begin{equation}
E^\mathrm{el}_0 =
T[\rho]+\int V_\mathrm{ext}(\mathbf{r}) \rho(\mathbf{r})\;d\mathbf{r}
+ E_\mathrm{H}[\rho]
+ E_\mathrm{xc}[\rho].
\end{equation}
The kinetic energy $T[\rho]$ is defined by
\begin{equation}
T[\rho]=
\sum_{i=1}^{N} -\frac{1}{2}
\int \psi^*_i(\mathbf{r})\Delta \psi_i(\mathbf{r})\;d\mathbf{r},
\end{equation}
which is not the true kinetic energy of the system of interacting electrons.
The Hartree energy $E_\mathrm{H}[\rho]$ is given by:
\begin{equation}
E_\mathrm{H}[\rho] =\frac{1}{2}
\int \int
\frac{\rho(\mathbf{r})\rho(\mathbf{r}')}{|\mathbf{r}'-\mathbf{r}|}\;d\mathbf{r}d\mathbf{r}'.
\end{equation}
DFT is exact in principle, however, the exchange-correlation functional $E_\mathrm{xc}[\rho]$ as well as $V_\mathrm{xc}[\rho](\mathbf{r})$ are not known and must be approximated.

We implement the variational method in \eqref{e1} to get the weak formulation of Kohn-Sham equation: Find $(\varepsilon_i, \psi_i) \in \mathbb{R}\times H^1_0(\mathbb{R}^3)$, $i=1, \cdots, N$, such that
\begin{equation}\label{e6}
   \left\{\begin{split}
      a( \rho; \psi_{i}, \varphi) = ( \varepsilon_{i} \psi_{i}, \varphi)\quad \forall \varphi\in H^1(\mathbb{R}^3),\\
     \int_{R^3}\psi_i(\mathbf{r})\psi_j(\mathbf{r})d\mathbf{r} = \delta_{ij}, i,j=1,2,\ldots,N,
   \end{split}
   \right.
\end{equation}
where
$$a( \rho; \psi, \varphi ) = \frac{1}{2}(\nabla\psi, \nabla\varphi)+(V_\mathrm{eff}(\rho)\psi, \varphi)\quad \forall \psi, \varphi\in H^1(\mathbb{R}^3).$$

From the density functional theory, we know that the ground state of the system can be obtained by
solving the lowest $N$ pairs of  eigenvalues and eigenvectors of the Kohn-Sham equation.

\subsection{Plane-wave discretization}
We now consider the periodic boundary conditions in a large volume $\Omega$ that is allowed to go to infinity.
In periodic solids, there is an infinite number of non-interacting electrons moving in an infinite external potential (such as the one generated by an infinite number of nuclei). However, Bloch's theorem \cite{Martin} can be invoked to express the wavefunction as the product of a cell-periodic part and wavelike part, whose wavevector is drawn from the first Brillouin zone (BZ) of the reciprocal lattice:
\begin{equation}
  \psi_{i,\mathbf{k}}(\mathbf{r})= u_{i,\mathbf{k}}(\mathbf{r})\mathrm{exp}[\mathrm{i} k\cdot \mathbf{r}],
\end{equation}
with $u_{i,\mathbf{k}}(\mathbf{r})=u_{i,\mathbf{k}}(\mathbf{r}+\mathbf{R})$ where $\mathbf{R}$ are the lattice vectors.

The infinite number of electrons in the solid is thus accounted for by an infinite number of $\mathbf{k}$ points in the BZ, and only a finite number of electronic states are occupied at each $\mathbf{k}$ point.
For instance, the electronic density is given by:
\begin{equation}
\rho(\mathbf{r})=\sum_{i=1}^{N_v} \int \left| \psi_{i,\mathbf{k}} (\mathbf{r}) \right|^2 \;d\mathbf{k},
\end{equation}
where $N_v$ is the number of occupied states.

Furthermore, the electronic wavefunctions at $\mathbf{k}$ points that are very close will be very similar.
Hence, it is possible to represent the electronic wavefunctions over a region of the BZ by the wavefunctions at a single $\mathbf{k}$ point.
This can be exploited for replacing integrals over the BZ by a weighted sum on a discrete mesh of well-chosen $\mathbf{k}$ points.
For instance, the one for the electronic density becomes:
\begin{equation}
\rho(\mathbf{r})=\sum_{i=1}^{N_v} \sum_{s=1}^{N_\mathbf{k}} w_s \; \left| \psi_{i,\mathbf{k}_s} (\mathbf{r}) \right|^2,
\end{equation}
where $w_s$ are the weights associated to the special $\mathbf{k}$-points $\mathbf{k}_s$ with $s=1, \ldots, N_\mathbf{k}$.
In the case of an homogeneous mesh, all the weights are equal and given by $w_s=\Omega_\mathrm{BZ} / N_\mathbf{k}$.
The accuracy of the calculations can always be increased by using a denser set of special $\mathbf{k}$-points.
In semiconductors, a modest number is sufficient to achieve a well-converged sampling density  because of the smoothly varying nature of KS states in $\mathbf{k}$-space.
In metals, however, much denser grids are required due to the abrupt change in the occupancy of each state with the wavevector $\mathbf{k}$.

The cell-periodic part of the wavefunctions can conveniently be represented as an expansion in terms of some set of mathematical basis functions.
The coefficients of the functions in this basis set are then the primary values used to build a computational representation.
Many different basis sets have been developed for use in periodic solid-state calculations (see Ref. \cite{Martin} for a detailed description).
The most natural (due to the periodicity) and popular (due to its ease of use) is the plane-wave basis set.
Each function $u_{i,\mathbf{k}}(\mathbf{r})$ is expressed as a Fourier series whose basis states are plane waves whose wavevector is a reciprocal lattice vector $\mathbf{G}$ (which are defined by $e^{i\mathbf{G} \cdot \mathbf{R}}=1$):
\begin{equation}
  u_{i,\mathbf{k}}(\mathbf{r})=\sum_\mathbf{G} c_{i,\mathbf{k}+\mathbf{G}} \times \frac{1}{\sqrt{\Omega}}\mathrm{exp}[\mathrm{i} \mathbf{G}\cdot \mathbf{r}].
\end{equation}

So a wave function can be written as
\begin{equation}\label{e2}
  \psi_{i,\mathbf{k}}(\mathbf{r})=\sum_\mathbf{G} c_{i,\mathbf{k}+\mathbf{G}} \times \frac{1}{\sqrt{\Omega}}\mathrm{exp}[\mathrm{i} (\mathbf{k}+\mathbf{G})\cdot \mathbf{r}],
\end{equation}
where $c_{i,\mathbf{k}+\mathbf{G}}$ are the expansion coefficients of the wave function.

Due to the fact that
 the coefficients $c_{i,\mathbf{k}+\mathbf{G}}$ for the plane waves with small kinetic energy $\frac{1}{2}|\mathbf{k}+\mathbf{G}|^2$
are typically more important than those with large kinetic energy \cite{Payne Teter Allan Arias Joannopoulos},  the plane-wave basis set can be truncated
to include only plane waves that have kinetic energies less than some particular cutoff energy $E_\mathrm{cut}$, i.e.
\begin{equation}\label{e3}
   \frac{1}{2}|\mathbf{k}+\mathbf{G}|^2 \leq E_\mathrm{cut}.
\end{equation}

The plane waves form an orthonormal basis set and do not depend on the location of the nuclei which simplifies the form of the equations and their implementation.
Furthermore, the size of the basis set (and therefore the accuracy of the calculations) can be systematically increased and easily controlled by a single parameter, the cut-off energy $E_\mathrm{cut}$ \cite{Hutter}, retaining only those $\mathbf{G}$-vectors such that $\frac{1}{2}|\mathbf{k}+\mathbf{G}|^2 \leq E_\mathrm{cut}$.
There are however two important disadvantages over other basis sets.
First, the number of basis functions required is quite large, which increases computational cost.
Second, it is quite difficult to represent sharp peaks in the KS states, such as those occurring in the core regions near nuclei due to the singularity of the electron-nuclear Coulomb attraction.

The states in the core region have however a negligible contribution to the
electronic properties of a material. Therefore, it is not necessary to represent them or the
Coulomb potential exactly. First, the states localized entirely within a core region,
called core states, may be precomputed (the frozen core approximation), avoiding the
need to include them explicitly in the calculation. Second, the Coulomb potential in
the core regions can be replaced with a pseudopotential which is constructed to
reproduce the atomic scattering properties and Coulombic form outside the core region, but
which is weaker and smoother inside. The remaining states, called valence states, are
described by pseudo-wavefunctions which are significantly smoother, hence improving the convergence with respect to $E_\mathrm{cut}$, without loss of accuracy \cite{CohenHeine, Phillips}.
The pseudopotential consists of two parts: one local part $V_\mathrm{loc}$, and a nonlocal part $V_\mathrm{nl}$. In the pseudopotential setting, the Kohn-Sham equation is  still formulated as \eqref{e1},  but $V_\mathrm{ext}(\mathbf{r})$ is  now being  $V_\mathrm{loc} (\mathbf{r}) + V_\mathrm{nl}(\mathbf{r})$, $N$ now being the
number of valence electrons, and $\{\psi\}_{i=1}^{N}$  being the set of the pseudo-wavefunctions  of the valence electrons.
The pseudo-wavefunctions can be approximated by far fewer basis functions \cite{Payne Teter Allan Arias Joannopoulos}. In this paper, we consider the pseudopotential case.

Therefore, we get a finite plane-wave discretization of \eqref{e6} as follows:
 \begin{equation}\label{e4}
   \left\{\begin{split}
     a( \rho;\psi_{n,i}, \varphi_{n'} ) = ( \varepsilon_{n,i} \psi_{n,i}, \varphi_{n'} ) \quad \forall \varphi_{n'}\in V_{N_G},\\
     \int_{R^3}\psi_{n,i}(\mathbf{r}) \psi_{n,j}(\mathbf{r})d\mathbf{r} = \delta_{ij}, i,j=1,2,\ldots,N,
   \end{split}
   \right.
\end{equation}
where $V_\mathrm{N_G}$ is a finite $N_G$ dimensional space spanned by the plane-wave basis, $N_G$ is the number of $\mathbf{G}$ satisfying (\ref{e3}), i.e
$$V_{N_G}  = \mathrm{span}\{\frac{1}{\sqrt{\Omega}}\mathrm{exp}[\mathrm{i} (\mathbf{k}+\mathbf{G})\cdot \mathbf{r}]\big|\mathbf{G}\in\mathbb{Z}, \frac{1}{2}|\mathbf{k}+\mathbf{G}|^2 \leq E_\mathrm{cut} \},$$
and
$$a( \rho;\psi_{n,i}, \varphi_{n'} ) = \frac{1}{2}(\nabla\psi_{n,i}, \nabla\varphi_{n'})+(V_\mathrm{eff}(\rho)\psi_{n,i}, \varphi_{n'})\quad \forall \psi_{n,i}, \varphi_{n'}\in V_{N_G}.$$

\subsection{Self consistent field iteration}
The Kohn-Sham equation is a nonlinear eigenvalue problem.
It is usually dealt with using a SCF approach \cite{Kresse and Furthmuller}.
Typically one starts with an initial guess for $\rho(\mathbf{r})$, then calculates the corresponding $V_\mathrm{eff}[\rho](\mathbf{r})$ and solves the Kohn-Sham equation for the $\psi_i(\mathbf{r})$.
From these one calculates a new density and starts again.
This procedure is then repeated until convergence is reached.
The following is the general algorithm of the self-consistent field iteration:
\begin{enumerate}
      \item Give initial input charge density $\rho_\mathrm{in}$.

      \item Compute the effective potential $V_\mathrm{eff}(\rho_\mathrm{in})$.

      \item Find $(\varepsilon_i,\psi_i) \in \mathbb{R}\times H_0^1(\mathbb{R}^3)$   satisfying
      \begin{equation*}
   \left\{\begin{split}
      a( \rho_{in}; \psi_{i}, \varphi) &= ( \varepsilon_{i} \psi_{i}, \varphi)\quad \forall \varphi\in H^1(\mathbb{R}^3),\\
     \int_{R^3}\psi_{i}(\mathbf{r}) \psi_{j}(\mathbf{r})d\mathbf{r} &= \delta_{ij}, i,j=1,2,\ldots,N.
   \end{split}
   \right.
\end{equation*}


      \item Compute the new output charge density $\rho_\mathrm{out}$.

      \item Convergence check: if not converged, use some  density mixing method to get the new input charge density $\rho_\mathrm{in}$, goto step 2; else, stop.
      \end{enumerate}

The variation of the charge density is often used as the criterion for the convergence of the self consistent field in the quantum chemistry calculation.
For the density mixing method in step 5, if we simply take $\rho_\mathrm{out}$ as the initial density of the next iteration, it converges too slowly or even does not converge. Therefore, it is very important to choose the proper density mixing method.
Many such density mixing  methods have been proposed so far.
The most widely used are the following ones: simple mixing \cite{Johnson}, Pulay's mixing \cite{Pulay,Pulay2}, Broyden's mixing method \cite{SinghKrakauerWang, Srivastava} and modified Broyden's mixing method \cite{Johnson, Kresse and Furthmuller}.
In this paper we use the modified Broyden's mixing method .

After plane-wave discretization and SCF iteration, we obtain the following large scale linear eigenvalue problem
 \begin{equation}\label{e5}
   H\Psi=\varepsilon S\Psi,
 \end{equation}
where $H=(h_{mn})_{N_G\times N_G}$ with
\begin{equation*}
  \begin{split}
h_{mn} &= a(\rho_{in}; \mathrm{exp}[\mathrm{i}(\mathbf{k}+\mathbf{G}_m)\mathbf{r}], \mathrm{exp}[\mathrm{i}(\mathbf{k}+\mathbf{G}_n)\mathbf{r}]) \\
&= \frac{1}{2}|\mathbf{k}+\mathbf{G}_m|^2\delta_{m,n}+ \frac{1}{\Omega}\int_{\Omega}V_\mathrm{eff}(\rho_{in})\mathrm{exp}[-\mathrm{i} (\mathbf{G}_m+\mathbf{G}_n)\cdot \mathbf{r}]\;d\mathbf{r}
  \end{split}
\end{equation*}
is the stiff matrix, $S$ is the overlap matrix. If we use the norm-conserving pseudopotentials, $S = I.$
 In tradition, people usually focus on solving the large scale linear eigenvalue problem repeatedly.
 However, the solution of  the large scale eigenvalue problem  requires   large scale orthogonal operation, which limits  large scale parallelization in supercomputer.

\section{Parallel orbital-updating approach}
Motivated by the good performance of the parallel orbital-updating approach in the real space method \cite{DGZZ}, we apply the similar idea to the reciprocal space setting so as to cure the poor parallel scalability of the traditional method in the reciprocal space. In fact, this is one of the series works on the parallel orbital-updating approach \cite{DGZZ}.
%
The following Algorithm 1  is the basic parallel orbital-updating algorithm for solving the Kohn-Sham equation based on plane-wave bases.\\


\begin{algorithm}[H]
\caption{(Basic parallel orbital-updating method)}
\begin{enumerate}
      \item Choose initial $E_{\mathrm{cut}}^{(0)}$ and then obtain $V_{N_G^{0}}$, give
      the initial data $(\varepsilon_i^{0},\psi_i^{0}), i=1, \cdots, N$, and let $n=0$.

      \item Increase $E_{\mathrm{cut}}^{(n)}$ to $E_{\mathrm{cut}}^{(n+1)}$ and then obtain $V_{N_G^{n+1}}$.

       \item For $i=1,2,\ldots,N$, find $\psi_i^{n+1/2}\in V_{N_G^{n+1}}$ satisfying $$a(\rho_\mathrm{in}^{n}; \psi_i^{n+1/2}, \varphi) = ( \varepsilon_i^{n}\psi_i^{n}, \varphi) \quad  \forall \varphi\in V_{N_G^{n+1}}$$ in parallel , where $\rho_\mathrm{in}^{n}$ is the input charge density obtained by the orbits obtained in the $n$-th iteration or the previous iterations.

      \item Find $(\varepsilon_i^{n+1},\psi_i^{n+1}) \in \mathbb{R}\times\tilde{V}_N$   satisfying

\begin{equation*}
   \left\{\begin{split}
     a(\rho_\mathrm{in}^{n+1/2}; \psi_i^{n+1}, \varphi) &= ( \varepsilon_i^{n+1}\psi_i^{n+1}, \varphi) \quad  \forall \varphi \in \tilde{V}_N,\\
     \int_{R^3}\psi_{i}^{n+1} \psi_{j}^{n+1} &= \delta_{ij}, i,j=1,2,\ldots,N,
   \end{split}
   \right.
\end{equation*}
      where $\tilde{V}_N = \mathrm{span}\{\psi_1^{n+1/2},\ldots,\psi_N^{n+1/2}\}$,  $\rho_\mathrm{in}^{n+1/2}$ is the input charge density obtained
      from $\psi_i^{n+1/2}$ and $\psi_i^{k}$ for $k\leq n$,  $i=1, \cdots, N$.

%
%

      \item Convergence check: if not converged, set $n=n+1$, go to step 2; else,  stop.
      \end{enumerate}
\end{algorithm}

Using algorithm 1, the solution of the large scale linear eigenvalue problem is replaced by the solution of a series of independent source problems and some small scale eigenvalue problems. Since the source problems are all independent, they can be solved in parallel intrinsically. For each source problem, we can use the traditional parallelization strategies, such as domain decomposition or parallelization in matrix vector multiplication. Therefore, our algorithm has  two level of which is advantageous for  large scale parallelization.  Besides,
   since the solution  of the source problems is much cheaper than that of eigenvalue problems, especially for large scale problems,
   our basic parallel orbital-updating algorithm  will reduce the computational cost.
   More features of this algorithm  are given in Ref. \cite{DGZZ}.
  It is worth noting that Algorithm 1 can be used staring from a small cutoff energy and then
    increasing it until the accuracy  is reached.

\section{Modified parallel orbital-updating approach}
%
As stated in Ref. \cite{DGZZ}, there are several options for each steps in Algorithm 1.
For example, one can calculate more orbitals if the initial guess is not good enough, or one can also choose  different source problems.

In this section, we will present two new modifications not mentioned in Ref. \cite{DGZZ}, which are denoted by Algorithm 2 and Algorithm 3 as follows.\\

\begin{algorithm}[H]
\caption{(Modified parallel orbital-updating method I)}
\begin{enumerate}
      \item Choose initial  $E_{\mathrm{cut}}^{(0)}$ and then get $V_{N_G^0}$, give 
       the initial data $(\varepsilon_i^0, \psi_i^0), i=1, \cdots, N$.  Let $n=0$.

      \item Increase $E_{\mathrm{cut}}^{(n)}$ to $E_{\mathrm{cut}}^{(n+1)}$, and obtain $V_{N_G^{n+1}}$.

      \item For $i=1,2,\ldots,N$, find $e_i^{n+1/2}\in V_{N_G^{n+1}}$ satisfying
    $$a(\rho_\mathrm{in}^{n}; e_i^{n+1/2}, \varphi) = -[(a(\rho_\mathrm{in}^{n}; \psi_i^{n}, \varphi) - \varepsilon_i^{n} (\psi_i^{n}, \varphi))]  $$
        in parallel , where $\rho_\mathrm{in}^{n}$ is the input charge density obtained by the orbits obtained in the $n$-th iteration or the former iterations.

      \item Find $\{\varepsilon_i^{n+1},\psi_i^{n+1}\} \in \mathbb{R}\times \tilde{V}_N$   satisfying
\begin{equation*}
   \left\{\begin{split}
     a(\tilde{\rho}; \psi_i^{n+1}, \varphi) &= ( \varepsilon_i^{n+1}\psi_i^{n+1}, \varphi) \quad  \forall \varphi \in \tilde{V}_N,\\
     \int_{R^3}\psi_{i}^{n+1} \psi_{j}^{n+1} &= \delta_{ij}, i,j=1,2,\ldots,N,
   \end{split}
   \right.
\end{equation*}
     %
      where $\tilde{V}_N = \mathrm{span}\{\psi_1^{n},\ldots,\psi_N^{n}, e_1^{n+1/2},\ldots,e_N^{n+1/2}\}$, $\tilde{\rho}(x)$ is the input charge density obtained from the previous orbits.

      \item Convergence check: if not converged, go to step 2; else, stop.
      \end{enumerate}
\end{algorithm}

For any $\psi \in V_{N_G}$, we define the projection operator $P_\psi: H^1_0(\mathbb{R}^3) \rightarrow \mathrm{span}\{\psi\}$ as:
\begin{eqnarray}
 P_\psi \varphi = (\psi, \varphi) \psi ~~~~~~~\forall \varphi \in H^1_0(\mathbb{R}^3),
\end{eqnarray}
then we can also define the following modified parallel orbital-updating algorithm.

\begin{algorithm}[H]
\caption{(Modified parallel orbital-updating method II)}
\begin{enumerate}
      \item  Choose initial  $E_{\mathrm{cut}}^{(0)}$ and then get $V_{N_G^0}$,  give the initial data $(\varepsilon_i^0, \psi_i^0), i=1, \cdots, N$.  Let $n=0$.

      \item Increase $E_{\mathrm{cut}}^{(n)}$ to $E_{\mathrm{cut}}^{(n+1)}$, and obtain $V_{N_G^{n+1}}$.

      \item For $i=1,2,\ldots,N$, find $e_i^{n+1/2}\in V_{N_G^{n+1}}$ satisfying
        \begin{equation*}
         (I - P_{\psi_i^{n}})(H - \lambda_i^n I)(I-P_{\psi_i^{n}})e_i^{n+1/2} = (\varepsilon_i^n I - H)\psi_i^{n}
       \end{equation*}
        in parallel, where  $\rho_\mathrm{in}^{n}$ is the input charge density obtained by the orbits obtained in the $n$-th iteration or the former iterations.

      \item Find $\{\varepsilon_i^{n+1},\psi_i^{n+1}\} \in \mathbb{R}\times \tilde{V}_N$   satisfying
      \begin{equation*}
   \left\{\begin{split}
     a(\tilde{\rho}; \psi_i^{n+1}, \varphi) &= ( \varepsilon_i^{n+1}\psi_i^{n+1}, \varphi) \quad  \forall \varphi \in \tilde{V}_N,\\
     \int_{R^3}\psi_{i}^{n+1} \psi_{j}^{n+1} &= \delta_{ij}, i,j=1,2,\ldots,N,
   \end{split}
   \right.
\end{equation*}
      where $\tilde{V}_N = \mathrm{span}\{\psi_1^{n},\ldots,\psi_N^{n}, e_1^{n+1/2},\ldots,e_N^{n+1/2}\}$, $\tilde{\rho}(x)$ is the input charge density obtained from the previous orbits.

      \item Convergence check: if not converged, go to step 2; else, stop.
      \end{enumerate}
\end{algorithm}

We can see that Algorithms 2 and  3 have all the features of Algorithm 1. The main difference is that the dimensions of the small scale eigenvalue problems are double of that in Algorithm 1.

\section{Numerical experiments}

In this section, we apply our parallel orbital-updating algorithms to simulate several crystalline systems: Si (silicon), MgO (magnesium oxide) and Al (aluminium) with different sizes to show the efficiency of our algorithms.
 Our algorithms are implemented in the software package  Quantum ESPRESSO  \cite{QE}, which is a mature and open-source computer codes for electronic-structure calculations and materials modeling at the nanoscale. It is based on density functional theory, it uses a plane-wave basis sets and pseudopotentials.

Currently Quantum ESPRESSO supports PAW (Projector-Augmented Wave) sets \cite{Blochl2}, Ultrasoft pseudopotentials \cite{Blochl, Vanderilt} and Norm-Conserving  pseudopotentials \cite{Hamann, TrMa}.
We use the Norm-Conserving pseudopotentials in our tests.
Quantum ESPRESSO  also provides various density mixing methods.
In our experiments, we choose the modified Broyden's mixing method.
There are some diagonalization methods in the Quantum ESPRESSO.
One is the Conjugate-Gradient-like band-by-band diagonalization, the other is the  Davidson iterative diagonalization.
Conjugate-Gradient-like band-by-band diagonalization (CG) uses less memory and is more robust compared to the Davidson iterative diagonalization with overlap matrix \cite{QE1}.
Therefore, we compare our our new algorithms with the Conjugate-Gradient-like band-by-band
diagonalization used in Quantum ESPRESSO.
In our tests, we set the convergence threshold for the density to $1\times10^{-7}$.
In the tests of our new algorithms, we did not use the possibility to gradually increase  the cutoff energy, that is, the cutoff energy is fixed.
For this special case, $V_{N_{G}^{0}}$ is not updated, and the step 3 and step 4 in all our algorithms  are carried out in
 $V_{N_{G}^{0}}$ repeatedly. All calculations are carried out on LSSC-III in the State Key Laboratory of
Scientific and Engineering Computing, Chinese Academy of Sciences,
and in part on the Ulysses computer cluster in SISSA.

We carefully checked that the total energies and the eigenvalues obtained by all our new methods converge to those obtained by the CG method
 if the latter converges. Indeed we should point out that we have also found some cases for which the CG method did not converge while our methods did,   as shown later.  Since the results of the modified parallel orbital-updating obtained by the Algorithm 3 are similar to those obtained by the Algorithm 2, we only list the results of the modified parallel orbital-updating methods obtained by the Algorithm 2 in the following numerical experiments.

We first introduce some notations which will be used in the following tables and figures.

$\bullet$ CG = Conjugate-Gradient-like band-by-band diagonalization method as implemented in Quantum ESPRESSO \footnote{The Conjugate-Gradient method  here is different from the Conjugate-Gradient method for the optimization problem and is mainly for solving the eigenvalue problem.}

$\bullet$ ParO = Basic parallel orbital-updating method (Algorithm 1)

$\bullet$ MParO = Modified parallel orbital-updating method I (Algorithm 2)



$\bullet$ $N_\mathrm{procs}$ = Number of processors

$\bullet$ $N_\mathrm{atom}$ = Number of atoms

$\bullet$ $N_\mathrm{band}$ = Number of orbitals

$\bullet$ $N_\mathrm{scf}$ = Number of SCF iterations

$\bullet$ $N_\mathrm{group}$ = Number of the groups of bands

As referred to in the end of Sections 3 and   4, our new algorithms are interesting for  large scale parallelization and for reducing the computational cost, especially for large scale system. This will be  illustrated in the following  numerical experiments.

\subsection{Good scalability of system size}

In this subsection,  two examples are used to show the advantages of our new algorithms in terms of their scaling as the system size increase.

\subsubsection{ MgO crystals}

The first test set consists of four MgO crystals made of $2\times2\times2$, $3\times3\times3$, $3\times3\times4$, and $4\times4\times4$ supercells, hence containing $32$, $108$, $144$, and $256$ magnesium and oxygen atoms, respectively.
All the crystals are sampled using the $\Gamma$ point only.
The cutoff energy is set to  $30 Ry$.
All results for these systems are obtained by performing the computation on one processor.

\begin{table}[!htbp]
\begin{center}
\scalebox{0.99}{\begin{tabular} {|r|r|r|r|r|r|r|r|r|} \hline
\multicolumn{3}{|c|}{} &\multicolumn{2}{|c|}{CG} &\multicolumn{2}{|c|}{ParO} &\multicolumn{2}{|c|}{MParO}   \\\hline
$N_\mathrm{atom}$ &$N_\mathrm{band}$ &DOFs &$N_\mathrm{scf}$  &Time (s) &$N_\mathrm{scf}$  &Time (s)&$N_\mathrm{scf}$  &Time (s) \\\hline
64   &128    & 6807    &10    &190  &28       & 392   &21        &285    \\\hline
216  &432   &23149     &12  & 5571  &20       &5456   &14        &5397   \\\hline
288  &576   &30063     &13    &13902   &21    &  12537   &14           &  12514          \\\hline
512  &1024  &54804     &12    &72109   &21    & 67407    &14           & 62825         \\\hline
\end{tabular}}
\caption{  Results  for MgO crystals of different sizes obtained  using CG, ParO and MParO with one processor.
 Both ParO and MParO are faster than CG, the advantage is more obvious for large system. }\label{t7}
\end{center}
\end{table}

\begin{figure}[H]
 \centering
 \includegraphics[width=10cm]{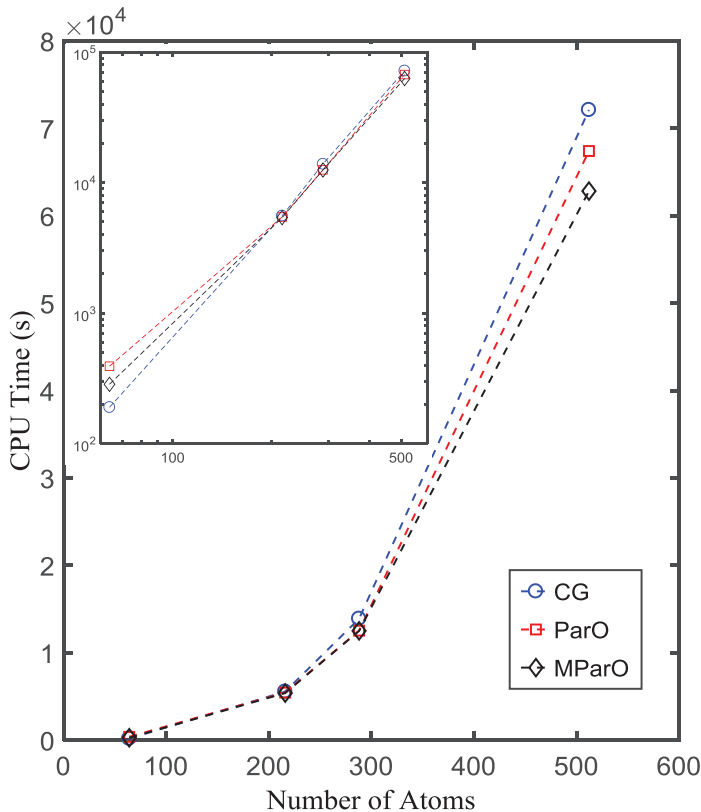}
 \caption{(Color online)  CPU time vs the system size (number of atoms) for the MgO crystals for the different methods:
 CG, ParO, and MParO are reported in blue, red, and black, respectively.  The larger the system, the more our algorithms are advantageous in  reducing the  computational cost.}
 \label{f5}
\end{figure}

Table \ref{t7} shows the detailed information for   MgO crystals  obtained by the different methods.
Fig. \ref{f5} shows the  CPU time as a function of   the system size for the different methods.
 From Table \ref{t7} it can be seen  that for small systems, the CPU time cost for our new methods is longer than that for  CG.
 However,  the CPU time cost for ParO and MParO increase slower than that for CG as a function of system size.
 From Fig. \ref{f5} we can see this more clearly, since the curves obtained by our methods are all below that obtained by  CG   as the system size increases.
The  log/log plot in the inset of  Fig. \ref{f5}  shows that the scaling of system size is similar for all the three methods. However, the original  plot in Fig. \ref{f5} shows that the pre-factors for ParO and MParO are smaller than that for CG.
 This shows that our methods reduce the computational cost compared to CG, especially for large systems.

\subsubsection{Aluminium crystals}

%

The second test set consists of two Al crystals of $3\times3\times3$ and $4\times4\times4$ supercells,
hence containing $108$ and $256$ aluminum atoms, respectively.
Generally, when dealing with a metal, a dense grid of k points should be used.
However, here, we are mainly interested in comparing the behavior of
the different methods for the same problem. Therefore, for simplicity, we use only
$\Gamma$-point sampling for both systems, and the kinetic-energy cutoff is set to
$30 Ry$. All results are obtained using one processor.

\begin{figure}[H]
\begin{minipage}[H]{0.50\linewidth}
    \centering{\includegraphics[width=6.0cm]{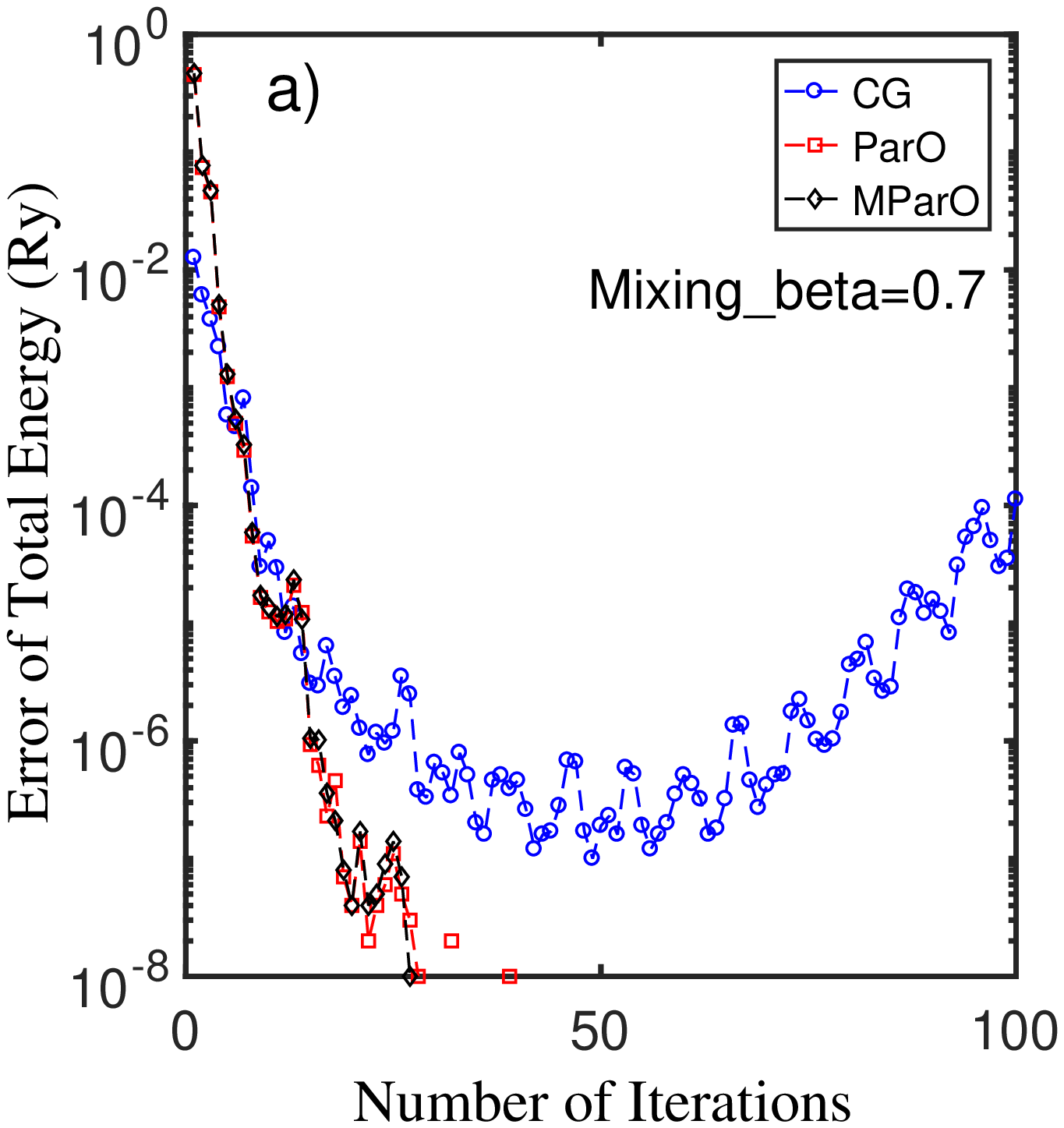}}
  \end{minipage}
  \hfill
  \begin{minipage}[H]{0.50\linewidth}
    \centering{\includegraphics[width=6.0cm]{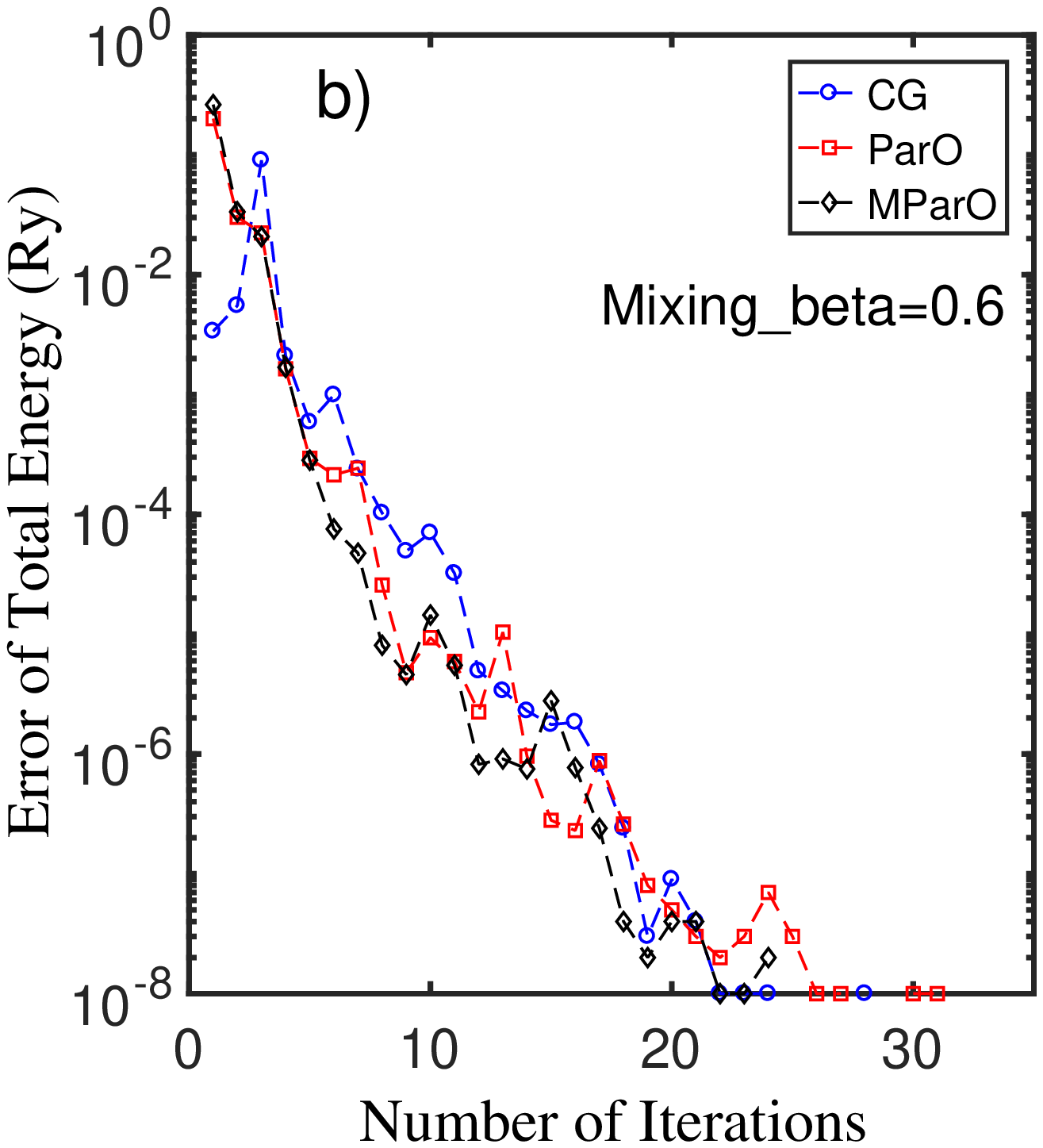}}
  \end{minipage}
  \caption{(Color online) The error of total energy vs the number of iterations for Al crystal made of $4\times 4 \times 4$ supercell  for the different methods for the default setting with different choice of mixing\_beta. CG, ParO, and MParO are reported in blue, red, and black, respectively.  When mixing\_beta is set to $0.7$, ParO and MParO  converge while CG  does not,  but when   mixing\_beta is set to $0.6$, all the three methods CG, ParO, and MParO converge.}\label{f6}
\end{figure}

Table \ref{t8} shows the detailed information for Al crystals obtained by the
different methods for the default setting where mixing\_beta parameter in the Broyden
mixing is set to $0.7$.  From Table \ref{t8} it can be seen that for the smaller system the total
energies by both our methods and CG converge. However, for the system which contains $256$
atoms, ParO and MParO converge while CG does not. This can be seen more clearly from Fig.\ref{f6}a,
 where the SCF error for Al crystal containing $256$ atoms as a function of SCF iteration by
the different methods is shown.

%
%

\begin{table}[!htbp]
 \centering
  \begin{tabular} {|c|c|c|c|c|c|c|c|c|c|} \hline
$N_\mathrm{atom}$        &DOFs  & Method                        &$N_\mathrm{scf}$              &Time (s)     &Error of energy        \\\hline
\multirow{3} {*}{108}  & \multirow{3} {*}{13805}      &CG               &    16      &   647   & $8\times 10^{-8}$            \\\cline{3-6}
                    &           &ParO                      &60           & 1534  & $5\times 10^{-8}$                \\\cline{3-6}
                 &            & MParO               &17           &570     &  $1\times 10^{-8}$           \\\hline
\multirow{3} {*}{256}                 & \multirow{3} {*}{37387}        & CG           & $*$         &   $*$        & $*$      \\\cline{3-6}
       &         & ParO                     &46           &  15917  & $5\times 10^{-8}$                \\\cline{3-6}
               &        &  MParO               &29           &  10239   & $1\times 10^{-8}$             \\\hline
   \end{tabular}
 \begin{tablenotes}
        \footnotesize
        \item[] $*$ For this case, we can not get the convergent results.
      \end{tablenotes}
 \caption{ Results for Al crystals of different sizes  obtained using CG, ParO and MParO with one processor for the default setting where mixing\_beta is set to $0.7$.  For large system, our methods ParO and MParO can get convergent results while CG can not.}\label{t8}
\end{table}

There are many strategies that can be adopted to improve SCF convergence and, for instance,
reducing mixing\_beta to $0.6$ is enough to make the CG method converge.
However our aim here is to compare the different methods in the same conditions. The results for
all  methods, CG, ParO, and MParO with the modified setting are reported in Table \ref{t8} and Fig. \ref{f6}b where it
can be seen that convergence, in terms of number of iterations needed to be achieved, is improved
for all methods, and ParO and MParO are competitive with or outperform CG in terms of timing.
Of course many more tests would be needed to draw general conclusions
about the relative merits of the different methods.

%

\begin{table}[!htbp]
 \centering
  \begin{tabular} {|c|c|c|c|c|c|c|c|c|c|} \hline
$N_\mathrm{atom}$        &DOFs  & Method                        &$N_\mathrm{scf}$              &Time (s)     &Error of energy            \\\hline
\multirow{3} {*}{256}                 & \multirow{3} {*}{37387}        & CG          &29           &  14222  & $1\times 10^{-8}$      \\\cline{3-6}
       &         & ParO                     &33           &  14580  & $3\times 10^{-8}$                  \\\cline{3-6}
               &        &  MParO              &25           &  10958   & $1\times 10^{-8}$            \\\hline
   \end{tabular}
  \caption{Results for an Al crystal made of $4\times 4\times 4$ supercell  obtained using CG, ParO and MParO with one processor for the case of mixing\_beta being set to 0.6. All the methods CG, ParO, and MParO converge.}\label{t9}
\end{table}

%

\subsection{Good scalability of parallelization}

In this subsection, we will use a Si crystal consisting of  a $5\times5\times5$ supercell with $1000$ silicon atoms as example to show the good parallel scalability of our new algorithms.   For this system, the number of computed orbitals is  $2000$.
The cutoff energies are set to be $45 Ry$ and the corresponding Brillouin zones are sampled by only the $\Gamma$-point.

%
%
%

\begin{figure}[H]
\centering
 \includegraphics[width=10cm]{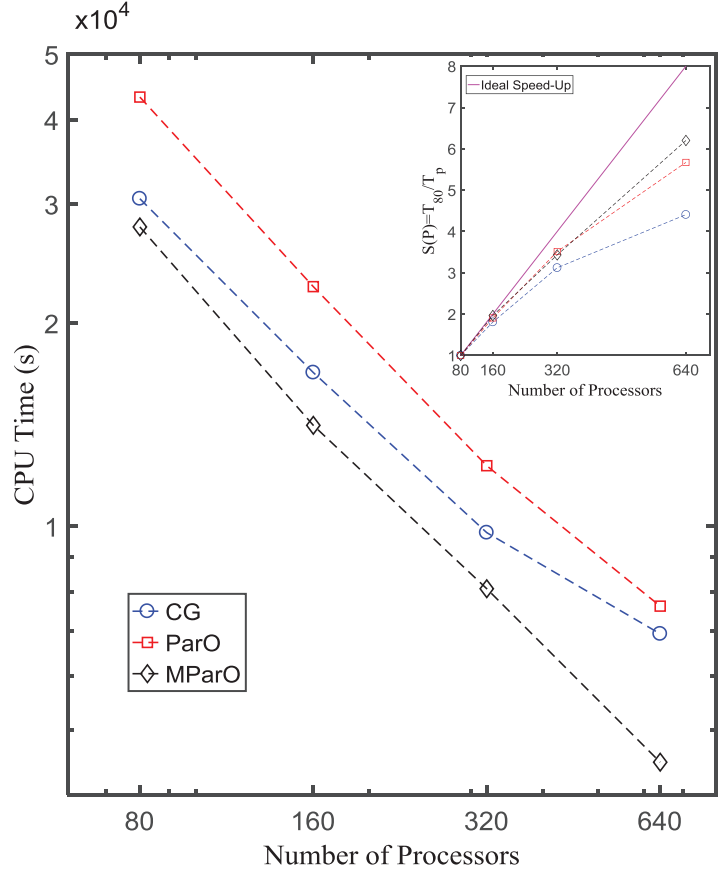}
  \caption{(Color online) CPU time vs the number of processors for a Si crystal with $1000$ atoms in the supercell  for the different methods:
  CG, ParO, and MParO are reported in blue, red, and black, respectively.   The inset shows the speed-up  obtained as $S(p) = \frac{T_{80}}{T_{p}}$ with $p$ the number of processors used, where $T_{p}$ is the wall time cost by $p$ processors, the purple line is the ideal speed-up $\frac{p}{80}$. The parallel scalability of ParO and MParO is better than that of CG.}\label{f4}
\end{figure}

Table   \ref{t11} show the detailed information for    Si crystal by the different methods  using $80$, $160$, $320$, $640$ processors, respectively.
Fig. \ref{f4} shows CPU time for  Si crystal as a function of   the number of processors for different methods.
For the system considered here,  it is known that  when the number of processors is smaller than $80$, the parallel efficiency of the plane-wave parallelization is relatively high.
Therefore,  for testing  our algorithms with $160$, $320$, $640$ processors,  the bands are divided into $2$, $4$, $8$ groups, respectively.
For each group,  $80$ processors are used for  the plane-wave parallelization.
For the CG method, since there is no band parallelization,
all processors are partitioned using only the plane-wave parallelization.

\begin{table}[!ht]
\begin{center}
\scalebox{1.0}{\begin{tabular}{|r|r|r|r|r|r|r|r|r|} \hline
   \multicolumn{1}{|c|}{} &\multicolumn{2}{|c|}{CG} &\multicolumn{3}{|c|}{ParO} &\multicolumn{3}{|c|}{MParO}   \\\hline
$N_\mathrm{procs}$ &$N_\mathrm{scf}$  &Time(s)&$N_\mathrm{group}$ &$N_\mathrm{scf}$  &Time(s)&$N_\mathrm{group}$&$N_\mathrm{scf}$  &Time(s) \\\hline
 80   &15     &30562 &1    &46   &43220  &1    &15          &27760     \\\hline
 160  &15     &16897   &2   &46    &  22647 &2   & 15           &14114  \\\hline
 320 &15      &9790  &4   &46        &12299 &4       & 15           &8086\\\hline
 640 &15      &6933  &8      &46       &7620 &8     & 15            & 4476\\\hline
   \end{tabular}}
\caption{ Results for a Si crystal with $1000$ atoms in the supercell obtained using CG, ParO and MParO with increasing number of processors.
The behavior of ParO and MParO is better and better with the increasing of processors, especially for MParO. }\label{t11}
\end{center}
\end{table}

 %


From Table \ref{t11}, it can be seen that  the CPU time cost for MParO is shorter than that for CG, while the CPU time cost for ParO is longer than that for CG.
However, from Fig. \ref{f4} we can see that when the number of processors is larger than  $320$,
the curves obtained by ParO and MParO  are steeper than that obtained by CG.
From this it can be seen  that the parallel scalability of our new methods is better than CG, especially for MParO.
To see it more clearly, one can also see the figure with speed-up  in the inset of  Fig. \ref{f4}. Since using $1$ processor can not obtain the
converged results for Si crystal with $1000$ atoms supercell due to the limitation of memory, the speed-up here is obtained by comparing the wall time for cases using different number of processors with that for case of using $80$ processors.
From the curves shown in Fig. \ref{f4}, the advantage of our methods in parallel scalability is obvious.

We should point out that, in our current tests, the cutoff energy is set to be a fixed value.
If we can start from a small cutoff energy and increase it until the convergence is reached, we can reduce the  computational cost further.
From this point of view, we believe our new methods will be more competitive than  CG.

\section{Discussion and conclusion}

Motivated by the parallel orbital-updating approach proposed in Ref. \cite{DGZZ,DLZZ},
 we propose some modified parallel orbital-updating methods  for the plane-wave discretization of the Kohn-Sham equation in this paper.
 We show that, by using the two-level parallelization of the orbital-updating approach, the poor parallel scalability of the plane-wave discretization can be largely improved.
Indeed our numerical experiments show that the parallel orbital-updating approach based plane-wave method has considerable  potential for carrying out  large system computation on modern supercomputers.

We should point out that our two-level parallelization only focuses on the solution of the associated eigenvalue problems resulting from the electronic structure calculations. In fact, in the electronic structure calculations, there are some other possibility for parallelization.   For example, when using hybrid functionals for approximating the exchange-correlation energy,  the  exchange potential can be obtained by solving many different Poisson equations, which can be done in parallel intrinsically. Any such kind of parallelization can be combined with our algorithms and
hence further increase the  parallelization.

As  we have pointed out at beginning and at the end of Section 5, the cutoff energy was set to  a fixed value in all our tests. To achieve the
gradual increase of the cutoff energy, one needs to design some efficient a posteriori error estimator to tell how to
evaluate and improve the approximate accuracy based on increasing the cutoff energy. It is indeed our on-going work to
design such kind of a posteriori error estimator and then increase the cutoff energy gradually until the accuracy has been reached,
 which will be addressed elsewhere.
 We believe that in that case, the parallel efficiency of our new algorithms will become even better.

\section*{Acknowledgements}
%
%
This work was partially supported by the National Science
       Foundation of China under grant 9133202, 11434004, and 11671389, the Funds for Creative Research Groups of
       China under grant 11321061,  the Key Research Program of Frontier Sciences and the National Center for Mathematics and Interdisciplinary
       Sciences of the Chinese Academy of Sciences, and the Fonds de la Recherche Scientifique (F.R.S.-FNRS), Belgium.

\end{document}